\documentclass[twocolumn]{revtex4-1}
\usepackage{graphicx}
\usepackage{dcolumn}
\usepackage{bm}
\usepackage{amssymb}
\usepackage{latexsym}
\usepackage{amsmath}
\newcommand{\be}{\begin{equation}}
\newcommand{\ee}{\end{equation}}
\newcommand{\bea}{\begin{eqnarray}}
\newcommand{\eea}{\end{eqnarray}}

\newcommand{\txi}{\tilde{\xi}}

\newcommand{\dtxi}{\dot{\tilde{\xi}}}
\newcommand{\bK}{\bar{K}}
\newcommand{\tK}{\tilde{K}}
\begin{document}
\title{A linear response relation in general relativity with applications to
dense matter relativistic stars }
\author{ Seema Satin \\
Indian Institute for Science Education and Research, Kolkata India \\
seemasatin@iiserkol.ac.in}
\begin{abstract}
A new formalism in general relativity with a linear response relation between
perturbed Einstein tensor and the stress-energy tensor is presented. Basic
 concepts
are borrowed from statistical physics and theory of stochastic processes
by extending them for a spacetime structure.  
 We show  qualitatively new results of the first applications which
lay foundations for a sub-hydro mesoscopic theory in dense matter relativistic
stars. This will enable one to probe structure and phenomena
at intermediate scales in relativistic stars and  exotic fluids that they are 
made up of. 
 The overall aim is to build foundations for studying dynamical equilibrium and 
non-equilibrium properties of astrophysical bodies based on 
 stochastic correlations of perturbations of spacetime and matter fields.
This will add to the present literature on asteroseismology a new insight in
a significant way. 
It is also expected that this will  lead to direct or indirect observational
 consequences
 of meso scale physics under extreme conditions  present in the exotic
 objects.
\end{abstract}
\maketitle
\textit{Introduction:}
 In this letter we introduce a  linear response relation in the
perturbed Einstein's equation which gives a new mathematical formalism with
 potential applications to asteroseimology and cosmology.
 The significance  lies in the fact that it
forms a theoretical base for studying mesoscopic scale physics in
general relativity. 
 This formalism essentially enables a non-local probe with two point
or higher order correlations obtained in a perturbative approach to analyse 
 structure, phenomena as well as dynamics inside a gravitating body. 
Such a motivation  arises from a general theoretical
point of view  inspired by other areas in physics like condensed matter,
statistical physics, fluid dynamics, complex systems etc.  It is 
expected that this will  eventually lead to  
 the search for observational consequences in astronomy   
 for sub-hydro meso-scale analysis,  
  when  spacetime curvature and general relativistic considerations
cannot be ignored. 

 The perturbed form of the Einstein's equations 
 $ \delta G^{ab}[h;x) =  8 \pi \delta T^{ab}[ h,\xi;x) $ on which
 asteroseismology is based is assumed heuristically.
The adiabatic perturbations of a relativistic star in terms of
 $h_{ab}(x) $ for the metric and 
 $\xi_a (x) $ the Lagrangian displacement vector for fluid trajectories, are
the fundamental degrees of freedom, in terms of which all the basic theoretical
results are worked out.   
In this letter using a new mathematical construct, we show how stochastic
 3-velocity perturbations, as a cumulative effect in the  exotic
 fluid, can drive the
metric perturbations at intermediate sub-hydro scales. We will also  work out
from the first principles, the expressions for 
correlations of the perturbations which form the base of dynamical equilibrium
and non-equilibrium studies in compact stars. 

The nature of dense matter in relativistic stars  is
 an active area of research from different perspectives \cite{1,2,3,4,5,6,7,8}. 
 Equation of state which connects the
 microscopic details with macroscopic properties \cite{9,10}, is not yet known
accurately for  exotic dense matter stars. In such objects the density of
 matter
 is higher than nuclear density and goes around and above $10^{17}$ kg /$m^3$.
 After
 core collapse, the newly born relativistic stars cool down and reach
a cold dense matter configuration usually modelled by a perfect fluid.
 The interiors are understood to be made up of a superfluid layer 
\cite{11,12,13} below the
 outer crust and interesting phenomena like pulsar glitches are
expected to originate due to pinning effects of the vortices \cite{14,15} in the
 superfluid. Such studies are recently coming up with an opening towards
 modelling mesoscopic physics in the dense superfluid matter. Two and three
fluid rheological models are also being studied in this regard to account for
 dissipation \cite{16}. The theoretical structure presented here has potential
to enhance and add to the present active directions in research 
\cite{21,22,23,24} which are being carried out by various groups. 

A  classical Einstein-Langevin equation has been recently proposed in
 \cite{17,18,19,20} in order to model
internal sources in the relativistic star which induce stochastic
 perturbations. 
  In this letter, we  give a new  formalism which is
 relevant for the case of external agencies perturbing the
relativistic star globally (like a binary companion or another astrophysical 
body passing by), this is qualitatively and mathematically ( as we show) 
different than the earlier proposal and in its domain of applications. 
The formalism presented here is likely to  find immediate and wider attention
among the astrophyics community, as well as progress for numerical simulations
will be desirable and accepted more readily as first step modification to
 existing  elaborate methods. 
 It is interesting to note that the dense matter stars give us an opportunity
 to study multi-scales in a strong gravity
setting which we cannot achieve in terrestrial laboratories. For example
the results with pinning effects of the quantum
vortices being responsible for pulsar glitches \cite{14,15}
 cannot be studied in terrestrial labs. 

\textit{The Formalism:}
The  ideas from  statistical physics which we extend to 
 the basics of general relativity are through  first principles  
 and conceptually new. The linear response theory is applicable to
 Hamiltonian systems, but
 it can have a very simple form for classical systems, without the need
to start with the Hamiltonian. The perturbations in spacetime
curvature and matter can be split into a system-environment 
setting and  then connected by a suitable response kernel,  
 phenomenologically as follows.
\be \label{eq:1}
 \delta G^{ab}[h;x) = \int \mathcal{K}[g;x-x') \delta T^{ab}[h,\xi;x') d^4x'
\ee
 where $\mathcal{K}[g;x-x')$  is the linear response kernel defined as a 
two point function on the background spacetime $g_{ab}$.
This introduces a  linear response relation in general relativity, where
the perturbations of the curvature tensor act as a linear response to 
the perturbations in the matter fields.
In general one can assume  the response kernel on a 
curved spacetime structure   of the form $\mathcal{K}[g; x,x')$,
however  the covariant conservation puts a 
restriction that it should have  $x-x'$ dependence.
This sets a limit that we can take the cumulative
effects of the matter field stress tensor in equation (\ref{eq:1}) over
 locally flat regions only. Fortunately
this is sufficent  for relativistic stars where a larger portion of the
interior at a depth  can be modelled easily in the flat spacetime 
approximation. 
One can see that formulating the
response kernel as a scalar (rather than a tensor), preserves many desirable
astrophysical conditions rather than complicating them unecessarily.  
This enables one to  define susceptibility  of the spacetime to the
 perturbations in matter fields represented by the response kernel, which is a
 new mathematical construct in relativity. 
We have to show that, since $\nabla_a \delta G^{ab}(x) =0 $ the rhs of the 
above equation is also covariantly conserved.  
\bea
 & & \mbox{To Prove:} \nabla_{a_x} \int \mathcal{K}(x-x') \delta T^{ab}(x')
 d^4x' = 0  \\
& & \mbox{The covariant derivative can then be applied as} \nonumber \\
& & \int \nabla_{a_x} \mathcal{K}(x-x') \delta T^{ab}(x') d^4 x' =0 
\label{eq:n1} 
\eea
Using the condition for the specific form of the response kernel
with $x-x'$ dependence, it can be  easily shown that
\be
  \nabla_{a_x} \mathcal{K}(x-x') = -\nabla_{a_{x'}} 
\mathcal{K}(x-x')   
\ee
Equation (\ref{eq:n1}) then takes the form,
\be
 -\int ( \nabla_{a_{x'}} \mathcal{K}(x-x'))
 \delta T^{ab}(x') d^4 x' =0 
\ee
\be
 \implies  -\int [\nabla_{a_x'}(\mathcal{K}(x-x') \delta T^{ab}(x')) 
- \mathcal{K}(x-x') \nabla_{a_x'} \delta T^{ab}(x')] d^4 x' 
\ee
Due to boundary conditions on the reponse kernel, the first term in the
 above is zero, giving,   
\be
 \int  \mathcal{K}(x-x') \nabla_{a_{x'}} \delta T^{ab}(x') d^4 x' =0 
\ee
To be consistent with the Bianchi identities in the unperturbed Einstein 
equation, we assume $\nabla_a \delta_x T^{ab}(x) =0 $ as necessary. Hence  
we see that the covariant conservation holds for equation (\ref{eq:1}). 
 For the mesoscopic scales that we intent to probe
  we assume $\mathcal{K}(x-x') \neq \delta^4(x-x') $. 
While for global macro hyrodynamic scales, this can be simply
 reduced to taking
$\mathcal{K}(x,x') = \delta^4(x,x')$, bringing  it to the regular 
form $ \delta G^{ab}(x) = 8 \pi \delta T^{ab}(x) $. 
One can ask the question, what is the need to introduce such a modified version
of the perturbed field equations as in this letter. The view we present,
  gives rise to a formalism which encodes intermediate scale details in a very
general way, that are not captured in the regular large scale macro
picture. It is  to be noted that the new proposal with a
response kernel  relation holds for perturbed 
Einstein's equations, while the unperturbed background form $G_{ab}(x) = 
T_{ab}(x) $ remains untouched.  
 Adopting this  way for formulating meso scale effects is well known in 
statistical physics and  other areas  like condensed matter. 
  Thus extending the ideas  to gravitating systems opens
up exciting new avenues in theory and  later for possible observations in
 astronomy. 
 
The perturbations of relativistic stars
 can be  (i) deterministic, which is closer to the
 well-known type, and (ii) stochastic, which have not been addressed so far 
in a noticeable way.
In this letter we focus on stochastic perturbations that can be 
induced by external agencies in a gravitating body. 
We have taken the linear response in the present formulation over the full 
stress tensor, while  in \cite{19,20} the linear response relation is
 introduced in
a mathematically different way. We propose that when an external agency drives
the perturbations, the form in this letter is physically more
appropriate. This significantly influences the theoretical formalism
and how its applications will proceed.  
One may ask, where does stochasticity arise from in the fluid ?
Such a situation can arise due to a gravitating neighbour 
  which induces turbulence or mechanically complex non-linear effects at 
multi-scales in the binary companion or by an astrophysical body passing by in
 the vicinity.
 These scales  are expected to be much above the microscopic or quantum scales
 and a little below the macroscopic hydro range. 
There can be  non-linear effects and other phenomena due to external
agencies, that
can give rise to stochasticity. 

 Proceeding with the formulations further, the solutions of equation
 (\ref{eq:1})  restricted to radial 
perturbations as the first exercise, can be obtained in closed analytical form
as follows. Solutions for rotating 
configurations and non-radial perturbations, for which one is likely to resort
 to numerical (semi-analytical) solutions are planned  for future  extensive
  work.  
 
   For the spherically symmetric star in Schwarzchild coordinates,
 the line element reads,

\be \label{eq:metric}
ds^2 = - e^{2 \nu(r)} dt^2 + e^{2 \lambda(r) } dr^2 + r^2 d \Omega^2
\ee
and  the matter fields for perfect fluid (describing cold dense matter
in hydrodynamic approximation) are given by,
\be
T_{ab} = (\epsilon+ p) u_a u_b + g_{ab} p
\ee
where the four-velocity in the static equilibrium reads, 
\be
u^a =  (e^{-\nu}, 0 , 0,0).
\ee

Radial perturbations to the static equilibrium can be introduced when there is
motion in the fluid such that it preserves the form (\ref{eq:metric}).
Introducing a radial 3-velocity in the fluid given by $v = e^{\lambda- \nu}
\dot{r} $, the field equations take the form
\bea
G^t_t = 8 \pi T^t_t & : & \nonumber \\
  e^{-2 \lambda} (\frac{1}{r^2} - \frac{2}{r}
\lambda ' )& - & \frac{1}{r^2} = - 8 \pi \epsilon \frac{1}{1-v^2} \label{eq:11}\\
G^r_r = 8 \pi T^r_r & : & \nonumber \\
e^{-2 \lambda}(\frac{1}{r^2} + \frac{2}{r} \nu' ) & - & \frac{1}{r^2} = 8
\pi ( \epsilon \frac{v^2}{1-v^2} + p ) \label{eq:2} \\
G^t_r = 8 \pi T^t_r & : & \nonumber \\
- \frac{2}{r} e^{-2 \nu} \dot{\lambda}  & = & 8 \pi e^{\lambda - \nu}
(\epsilon + p) \frac{v}{1- v^2} \label{eq:3}\\
e^{ 2 \lambda} G^\theta_\theta = 8 \pi e^{2 \lambda}  T^\theta_\theta
& : & \nonumber \\
\nu '' + \nu'^2 & - & \nu ' \lambda ' +  \frac{1}{r} (\nu ' - \lambda ' )
= 8 \pi e^{2 \lambda} p
\eea
It can be checked that, on putting $v=0$ in the above equations, we get the
 static configuration easily. 
  While the non-radial perturbations are important from
 GWs point of view, the radial perturbations have their significance
to study stability properties. 
The radial velocity perturbations can be introduced, such that
\be \label{eq:perb}
\delta v = e^{\lambda - \nu} \dot{\xi}
\ee
where $\xi \equiv \xi_r$  for the radial component, is the only non-zero fluid
 displacement vector and "." denotes derivative w.r.t time.

The adiabatic fluid  perturbations (of the fluid variables) are given (in 
the standard form \cite{Friedman}) by
\bea
\delta u^a & = & q^a_b \mathcal{L}_u \xi^b + \frac{1}{2} u^a u^c u^d h_{cd}
\mbox{ (where $q^a_b = u^a u_b + \delta^a_b $) }  \nonumber \\
\delta \epsilon & = &  - \frac{1}{2} ( \epsilon + p) q^{ab}( h_{ab} + \nabla_a
\xi_b + \nabla_b \xi_a ) - \bm{\xi}\cdot \nabla \epsilon \nonumber \\
\delta p & = & -  \frac{1}{2} \Gamma_1 p q^{ab}( h_{ab} + \nabla_a
\xi_b + \nabla_b \xi_a ) - \bm{\xi}\cdot \nabla p
\eea
where $\Gamma_1 = \frac{\epsilon+ p}{p} \frac{dp}{d\epsilon}$ is the
 adiabatic index. For the radial perturbations, the above reduce to 
\bea
\delta p(r,t) & = &
- \Gamma_1 p \frac{e^{-\lambda}}{r^2} \delta \lambda - \Gamma_1 p
\frac{e^{-\lambda}}{r^2} [ e^{\lambda} r^2 \xi ]' - \xi p' \\
\delta \epsilon (r,t) & = &
 - (p + \epsilon) \frac{e^{-\lambda(r)}}{r^2} \delta \lambda(r)
- (p + \epsilon) \frac{ e^{-\lambda}}{r^2} [ e^{\lambda} r^2 \xi]'
- \xi \epsilon' \\
\delta u^r(r,t) & = &  \mathcal{L}_u \xi^r = e^{-\nu} \dot{\xi}
\eea
To solve the perturbed Einstein's equations with the response relation
 we will need only two components of equation (\ref{eq:1}), 
\begin{widetext}
\bea
\delta G^t_r(x)= \int \mathcal{K}(x-x') \delta T^t_r(x') d^4 x' & : &
\nonumber \\
- \frac{2}{r}e^{- 2 \nu} \dot{( \delta \lambda )}(r,t) & = & 8 \pi
 \int K(t-t') e^{2(\lambda - \nu ) } (\epsilon + p) \dot{\xi}(t',r) dt'
\label{eq:12}  \\
\delta G^r_r(x) = \int \mathcal{K}(x-x') \delta T^r_r (x') d^4 x' & : &
 \nonumber \\
\delta \nu'(r,t) & = & - e^{2(\lambda- \nu)} ( \frac{1}{r} + 2 \nu') \delta
\lambda' - 4 \pi r e^{2 \nu} \Gamma_1 p \int \delta \lambda(r,t') K(t-t') dt'
\nonumber \\
 + 4 \pi r e^{2 \nu} \Gamma_1 p \int K(t-t') [ \xi'(r,t')
+ (\frac{2}{r} + \lambda' + \frac{\epsilon'}{\epsilon+ p}) \xi(r,t') ] dt'
\label{eq:13}
\eea
\end{widetext}
Where we  consider the response kernel having an explicit form  $
\mathcal{K}(x-x') = K(t-t') \delta(r-r') \delta(\theta - \theta')
\delta(\phi - \phi') $.
Other cases can be taken into considered for modelling specific types of
 responses. For example,  
  $\mathcal{K} = K(t-t',r-r') \delta(\theta- \theta')
\delta(\phi- \phi') $ for the spherically symmetric case, where the cumulative
effect can be taken over radial depths and temporal coordinate. 
However for mathematical simplicity here, we take the earlier case.
We  also assume the Lagrangian displacement  of the form 
$\xi(r,t) \equiv \txi(r,t) e^{\gamma_r t}$ where the amplitude $\txi(r,t)$ is
stochastic, and $\gamma_r$ is the complex frequency at radial depth '$r$'. 
With  details of the calculations given in  the appendix, 
and presenting the main results here, by 
using the relation $ \xi(r,t) = e^{\nu - \lambda} \int \delta v(r,t') dt' $
 from (\ref{eq:perb}) where stochastic velocity perturbations
 $\delta v$ with a distribution $P(\delta v)$ are substituted, the following
 can be obtained easily,
\begin{widetext}
\bea
\delta \lambda(r,t) & = & (\nu' + \lambda')
\int \{ \bK(\gamma_r) e^{\nu- \lambda} - \tK(\gamma_r)( e^{\nu- \lambda}
 \delta(t-t')- \gamma_r) \} \delta v (r,t') dt' \label{eq:v1} \\
\mbox{ and } \nonumber \\
\delta \nu(r,t) & = & \int \int F_1(r',t') \delta v(r',t') dt' dr' 
\label{eq:v2}
\eea
where
\bea
F_1(r',t') & = & a_1(r) + \{ \delta(t-t') a_3(r') - a_3'(r') \} e^{\nu -
\lambda} + (e^{\nu - \lambda} \delta(t-t') - \gamma_r ) \{ a_2(r') \nonumber\\
& &  + a_4 (r') \delta(r-r') - a_4'(r') \}
\eea
\end{widetext}
The relations for $a_1, a_2, a_3, a_4$ and for $\bar{K}(\gamma_r), \tilde{K}(
\gamma_r) $  which denote susceptibility of spacetime, are given in the
 appendix.
Given that $\delta v$ has a stochastic distribution $P(\delta v)$, $\delta 
\lambda$ and $\delta \nu$ are also stochastic in nature. 
We may theoretically consider the velocity to have a Gaussian  distribution 
with $ \langle \delta v (r,t) \rangle = 0 $ as the first possible case, 
where $\langle ... \rangle $ denotes statistical average, giving the result, 
$\langle \delta \lambda(r,t) \rangle =0$ and $\langle \delta \nu(r,t) \rangle
 =0 $ from equations (\ref{eq:v1}) and (\ref{eq:v2}). 
A more interesting case to consider is that of the  Gaussian distribution with
 a non-zero mean.
 In the case of Gaussian distribution the one and two point expectation values
suffice to know the statistical properties of the system. 
In general taking  non-gaussian distributions for the velocity
 fluctuations as in a turbulent flow can be interesting for
two or three -component superfluid  interiors. 
 The stochastic perturbations  have meaning only as statistical averages, 
whether they are Gaussian or non-Gaussian. 

\textit{Results:}
The two point correlations of the metric perturbations  given above 
 take the form, 
\begin{widetext}
\bea
 \langle \delta \lambda^*(r_1,t_1) \delta \lambda(r_2,t_2) \rangle & = &
(\nu'(r_1) + \lambda'(r_1) ) (\nu'(r_2) + \lambda'(r_2))
e^{(\nu(r_1) + \nu(r_2) - \lambda(r_1) - \lambda(r_2) )} e^{\gamma^*_{r_1}t_1
+ \gamma_{r_2} t_2 } \int  \{\bK(\gamma_{r_1}) - \nonumber \\
& &  \tK(\gamma_{r_1}) \delta(t_1 - t_1') \} \{ \bK(\gamma_{r_2}) -
\tK( \gamma_{r_2}) \delta(t_2 -t_2') \} 
\langle \delta_s v^*(r_1,t_1') \delta_s v(r_2,t_2') \rangle dt_1' dt_2' 
\label{eq:lambdalambda}\\
\langle \delta \nu^*(r_1,t_1) \delta \nu(r_2,t_2) \rangle & = &   \int
F_1(r_1',t_1') F_1(r_2',t_2')\langle \delta v^*(r_1',t_1') \delta v(r_2',t_2')
\rangle dt_1'dt_2' dr_1' dr_2' \label{eq:nunu}  \\
\langle \delta \lambda^*(r_1,t_1)  \delta \nu(r_2,t_2) \rangle & = &
 (\nu'(r_1)+ \lambda'(r_1))\int [ \{\bK(\gamma_{r_1})e^{\nu(r_1) -
 \lambda(r_1) }- \tK(\gamma_{r_1}) ( e^{\nu(r_1) -\lambda(r_1)}
\delta(t_1 - t_1') - \gamma_{r_1} ) \} F_1(r_2',t_2') \nonumber \\
& & 
\langle \delta_s v^*(r_1,t_1') \delta_s v(r_2,t_2') \rangle dt_1'
 dt_2' dr_2' \label{eq:lambdanu}  \\
\langle  \delta \nu^*(r_1,t_1) \delta \lambda(r_2,t_2)\rangle & = &
(\nu'(r_2) + \lambda'(r_2)) \int F_1(r_1',t_1')[\{\bK(\gamma_{r_2})
e^{ \nu(r_2) - \lambda(r_2) } - \tK(\gamma_{r_2}) ( e^{\nu(r_2)- \lambda(r_2)}
\delta(t_2 - t_2') - \gamma_{r_2}) \}] \nonumber \\
& &  \langle \delta_s v^*(r_1,t_1') \delta_s v(r_2,t_2') \rangle dt_1'
dt_2' dt_r' \label{eq:nulambda}
\eea
\end{widetext} 
Similarly one can obtain 3 point and higher order correlations for non-
Gaussian distributions. 
We see that the correlations of the metric  perturbations are related 
to and integrated over  the
 correlations of the 3-velocity perturbations. 
It is important to note here that these are solutions of homogenous
form of perturbed Einstein's equations as opposed to the Einstein-Langevin
 equations
which are inhomogenous set of equations with a source term. Hence these 
results here are  different than in \cite{19,20}. The two formalisms are
mathematically, physically as well as in the  domain of applications
different, though both address stochastic perturbations in astrophysics.  

In the above expressions the interval between the two
spacetime points can be taken to be arbitrary, such that, depending upon our
 interest we can place them and probe the regions we wish to analyse.
This can be useful for numerical simulations of the interior structure of
the relativistic stars. We can
 consider  cases of two different radial points $r_1$ and
 $r_2$, at a given time $t$, or at two different 
times $t_1$ and $t_2$ . The other case could be that, of the same radial point,
 at two different times 
$t_1$ and $t_2 $. With the above expressions one can also get the root mean
 square of the perturbations, by taking the coincident limit. 
The 
separation in the radial/spatial distance of the points has to be consistent
 with the causality condition. These correlations form the basic building
 blocks for dynamical equilibrium and non-equilibrium studies in general
 relativity in order to  probe extended structure inside the massive
 astrophysical objects. 
It is well established in asteroseismology that the
 perturbations of the potentials from the interiors of relativistic stars
 can  travel through and
 out of the configuration while growing (or decaying) in amplitude
 significantly. 
 It is expected that near phase changes in the matter due to
 gravitational instability, these two point  or higher correlations may become 
  significantly large for specific intervals or separations. The evolution of
 the correlations then can be useful
 to investigate the extended non-local dynamics  of the compact object.
 The mesoscopic characteristics are expected to be carried by such 
stochastic correlations combined with the following perturbed Euler equations,
  $\delta \nabla_a T^a_1(x) =0 $,  which takes  the form
\begin{widetext}
\bea \label{eq:xiform}
& & e^{2(\lambda - \nu )} (\epsilon + p) \ddot \txi(r,t) +
 \{ e^{2(\lambda - \nu) } (\epsilon + p) \gamma_r + (\Gamma_1 p' - \nu')
\tK(\gamma_r) + p(r) a_2(r) \} \dot \txi(r,t)  \nonumber \\
& & + p \{ \Gamma_1 \tK(\gamma_r) + a_4(r) \} \dtxi'(r,t) -
\{\Gamma_1 p [(\nu' + \lambda') \bK(\gamma_r) + (\lambda' + \frac{2}{r}) -
\nu' ] + p'(1+ \Gamma_1) \} \txi' - \Gamma_1 p \txi''(r,t) + \nonumber \\
& &
[ e^{2(\lambda - \nu)} (\epsilon+ p) \gamma_r^2 - \Gamma_1 p'
\{(\nu' + \lambda') \bK(\gamma_r) + (\lambda' + \frac{2}{r}) \}
- \Gamma_1 p \{ \bK(\gamma_r) ( ( \nu'' + \lambda'') + \nu'(\nu'+ \lambda'))
 +  (\lambda'' - \frac{2}{r^2}) \nonumber \\
& &  + (\lambda' + \frac{2}{r} ) \nu' \} - p'' + \nu' p' ] \txi(r,t) = 0
\eea
\end{widetext}
It is this equation that can be used for stochastic mode analysis. However, we 
see that the complexity of the above differential equation will need further
 investigations into the methods of solution.
 One can notice the presence of
the term with $\dtxi(r,t)$ here, which can be associated with dissipation-like
 effect. It is the stochastic
 nature of the amplitude of the oscillations that gives rise to this extra 
term with $\dtxi(r,t)$ and also adds another one $\dtxi'(r,t)$.
 This form of the Euler equation is  new and touches upon the
fundamentals of the theory. It shows how stochastic perturbations in
 asteroseismolgy should evolve. This equation can  be combined with the 
perturbed equation of state, as a trial to obtain numerical solutions. We
 leave this as future directions and conclude this letter with the above
 new set of equations from (\ref{eq:lambdalambda}) to (\ref{eq:xiform})
as the opening results of an elaborate theory yet to be fully formulated.

\textit{Conclusion:}
 Modelling turbulence in the superfluid \cite{nils1,Aur,nils2,nils3,nils4} of
 the
relativistic stars is an active and open area of research, with which we
expect to relate one of our directions of research  in upcoming
 articles.
 It is well established that turbulence 
is connected to velocity field fluctuations at multi-scales from large to small eddies. For directly addressing eddies in the superfluid, one needs to consider
a rotating relativistic star configuration, which will need numerical 
solutions on the lines of work done here. 
 Developing such a mesoscopic theory will shed light on a vast 
range of scales of the interiors of dense matter stars and improve our
understanding about how the micro-scales are connected to macro scale physics
 in the exotic matter. Different forms of response kernel and distribution of 
fluid perturbations can address various possibilities that arise inside
the compact matter.

With  this letter, we open up an area for theoretical foundations
 in general relativity in order to study 
sub-hydro  mesoscopic physics in dense compact matter, relativistic fluids and
cosmology.  

\textit{Data availability statement :} No data was created in this work. 

\textit{Acknowledgements:}
This work has been funded by DST India, through grant number
 DST/WoS-A/PM-3/2021. The author is thankful to IISER Pune, India for providing
library facilities on an extended period of visit when major part of this letter
was finalized. 

\begin{center}
\textit{Appendix : }  \label{sec:appendix}
\end{center}

Proceeding with the solutions, from equation (\ref{eq:12}) we 
 obtain
\bea
\delta \lambda(r,t)& =& (\nu' + \lambda') [ \txi(r,t) \bK(\gamma_r) -
\dtxi(r,t) \tK(\gamma_r) ] e^{\gamma_r t}
\eea
 after using an ordinary Taylor series expansion ( since we
consider the region when a flat spacetime approximation is valid
locally)  for $\tilde{\xi}(r,t-\tau)$ around '$t$'. 

We have used for the term  $\int K(t-t') \xi(r,t') dt' $, 
\be
 \int K(\tau) \xi(r, t-\tau) d\tau  \mbox{ where } t-t'= \tau
\ee
 with the Taylor expansion  for $\tilde{\xi}(r,t-\tau)$, upto first order
\bea
& & \int K(\tau) \{\txi(r,t) + \tau \dtxi(r,t) \} e^{ \gamma_r (t-\tau)}
 d\tau \} \nonumber \\
\eea
 taking the Laplace transform is also possible and no 
mathematical issues arise in flat spacetime approximation. Thus,
\bea
& &  \int K(\tau) e^{-\gamma_r \tau} d\tau = \bK(\gamma_r)
\mbox{ similarly if we put $K(\tau) \tau = K_1(\tau)$, then ,}  \nonumber \\
& & \int K_1(\tau)
 e^{-\gamma_r \tau}  = \tK(\gamma_r)  \nonumber \\
\eea
from (\ref{eq:13}) we get,
\begin{widetext}
\bea
\delta \nu(r,t) & = & \int [ a_1(r) \txi(r,t) e^{\gamma_r t} + a_2(r)
\dtxi(r,t)e^{\gamma_r t}  + a_3 (r) \txi'(r,t) e^{\gamma_r t} + a_4(r)
\dtxi'(r,t) e^{\gamma_r t } ]dr'
\eea
where
\bea
a_1(r) & = & - e^{-2(\lambda - \nu)}( \frac{1}{r} + 2 \nu') (\nu'' +
\lambda'') + 4 \pi r e^{2 \nu} \Gamma_1 p  \bar{K}^2(\gamma_r) \{(\nu' +
\lambda') \bK(\gamma_r) - (\frac{2}{r} + \lambda' + \frac{\epsilon'}{
\epsilon+p})\}  \nonumber \\
a_2(r) & = & e^{-2 \nu} \tK(\gamma_r) \{ e^{- 2 \lambda}( \frac{1}{r} - 2 \nu'
 )(\nu'' + \lambda'') + 4 \pi r \Gamma_1 p(\frac{2}{r} + \lambda' +
\frac{\epsilon'}{\epsilon + p}) \} \nonumber \\
a_3(r) & = & - \bK(\gamma_r) \{ e^{-2(\lambda - \nu)} (\frac{1}{r} - 2 \nu')
 (\nu' + \lambda') + 4 \pi r e^{2 \nu} \Gamma_1 p \} \nonumber \\
a_4(r) & = & \tK(\gamma_r) \{e^{-2(\lambda - \nu)}(\frac{1}{r} + 2 \nu')  +
4 \pi r e^{2 \nu} \Gamma_1 p \}
\eea
\end{widetext}

\begin{thebibliography}{00}
\bibitem{1}J.M.Lattimer and M.Prakash, ApJ 550, 426 (2001).
\bibitem{2}J.L.Zdunik, M.Fortin and P.Haensel, A\&A , 599 (2017).
\bibitem{3}Jianing Li et. al ., Phys Rev D , 106,083021 (2022)
\bibitem{4} Golam Mortuza Hossain, Susobhan Mandal, Phys. Rev D 104, 123005
 (2021)
\bibitem{5}T.Celora, et.al, Phys.Rev D , 103016 (2022).
\bibitem{6} Tuhin Malik et.at., The Astrophys. J, 930:17 (15pp) (2022)
\bibitem{7} Kip S.Thorne and Alfonso Campolattaro., The Astrophys.J, 149 (1967).
\bibitem{8} Kostas D.Kokkotas and Bernd. G.Schmidt., Liv.Rev.Rel, 2(2) (1999).
\bibitem{9} Andreas Schmitt., Lect.Notes. Phys. 811, Springer, Berlin
 Heidelberg (2010).
\bibitem{10} Schmitt.A, Shternin.P, Astro. Sp.ScoLib, 457 (2018)
\bibitem{11} Gordon Baym et.al., Nature 224, 673-674 (1969).
\bibitem{12} Haskell,B.,Dedrakian,A. Astrophysic and Space Science Library,
 vol 457. pp 401-454, Springer, Cham. (2018).
\bibitem{13} Nils Andersson., Universe 7(1) (2021).
\bibitem{14} S.Seveso et.al, MNRAS, 455(4) (2016).
\bibitem{15} P.M.Pizzochero et.al, Nature Astronomy 1 (0134) (2017)
\bibitem{16} L. Gavassino., Class. Quant. Grav. 40 (16) 165008 (2023).
\bibitem{21} Chun Yuen Tsang et. al, Nature Astronomy 8, 328-336 (2024).
\bibitem{22} N.Andersson, T.Sidery,G.L Comer., MNRAS, 381 (2) (2007)
\bibitem{23} David Radice., ApJL 838 (1) (2017).
\bibitem{24} T.Celora et.al, Phys. Rev D 104, 084090 (2021).
\bibitem{17} Seema Satin., Gen.Rel.Grav 50:97 (2018).
\bibitem{18} Seema Satin., Gen. Rel.Grav 51(4) (2019).
\bibitem{19} Seema Satin., Class. Quant.Grav. 40 (5) (2023)
\bibitem{20} Seema Satin,. Gen. Rel. Grav. 55(37) (2023).
\bibitem{Friedman} John L. Friedman and Nikolaos Stergioulas., Rotating
Relativistic stars, Cambridge University Press. (2013).
\bibitem{nils1} N.Andersson, T.Sidery, G.L Comer., MNRAS 381(2) (2007).
\bibitem{Aur} Aurelien Sourie et. al., Phys Rev D 93, 083004 (2016).
\bibitem{nils2} N Andersson et.al, Class.Quantum Grav 37 085014 (2020).
\bibitem{nils3} Thomas Celora et.al, Phys.Rev D 104, 084090 (2021).
\bibitem{nils4} Thomas Celora et. al, Phys.Rev D 110 (12), 123039 (2024).
\end{thebibliography}
\end{document}